\documentclass[12pt]{article}

\newcommand{\bbr}{I\!\! R}
\newcommand{\bbz}{Z\!\!\! Z}

\newcommand{\3}{$^3$}

\newcommand{\x}{arXiv:}
\newcommand{\m}{\mathrm}

\usepackage{graphicx}

\begin{document}
\thispagestyle{empty}
\begin{center}

\null \vskip-1truecm \vskip2truecm

{\Large{\bf \textsf{Black Hole Final State Conspiracies}}}

{\large{\bf \textsf{}}}

{\large{\bf \textsf{}}}

\vskip1truecm

{\large \textsf{Brett McInnes}}

\vskip1truecm

\textsf{\\  National
  University of Singapore}

\textsf{email: matmcinn@nus.edu.sg}\\

\end{center}
\vskip1truecm \centerline{\textsf{ABSTRACT}} \baselineskip=15pt
\medskip

The principle that unitarity must be preserved in all processes, no
matter how exotic, has led to deep insights into boundary conditions
in cosmology and black hole theory. In the case of black hole
evaporation, Horowitz and Maldacena were led to propose that
unitarity preservation can be understood in terms of a restriction
imposed on the wave function at the singularity. Gottesman and
Preskill showed that this natural idea only works if one postulates
the presence of ``conspiracies" between systems just inside the
event horizon and states at much later times, near the singularity.
We argue that \emph{some} AdS black holes have unusual internal
thermodynamics, and that this may permit the required
``conspiracies" if real black holes are described by some kind of
sum over all AdS black holes having the same entropy.

\newpage

\addtocounter{section}{1}
\section* {\large{\textsf{1. Preserving Unitarity}}}
The least-understood regions of spacetime are its spacelike
``edges": the beginning of time and its end, whether the end be in
Crunches [the end state of a Coleman-De Luccia bubble
\cite{kn:deluccia} which nucleates with negative vacuum energy] or
inside a black hole. Without an understanding of the laws governing
these regions, we cannot arrive at a complete account of our
observations regarding the remainder of spacetime.

The principle of \emph{unitarity} \cite{kn:hsu} is one of our most
powerful tools for probing these regions. For example, Carroll and
Chen \cite{kn:carroll} used it to argue that Inflation \emph{alone}
cannot explain the very unusual conditions which obtained at the
earliest times; instead, one needs a theory of inflationary initial
conditions \cite{kn:albrecht}\cite{kn:couleinitial}. Similarly, it
has been argued convincingly \cite{kn:wittenconf}\cite{kn:juan} that
the AdS/CFT correspondence indicates that unitarity is [somehow]
preserved during black hole evaporation. If we can understand how
this works, we can hope to probe the final state of a black hole
interior.

Horowitz and Maldacena \cite{kn:horomald} attempted to obtain a
detailed understanding and implementation of this last idea by
making a proposal for the final quantum state inside a black hole.
The projection onto this unique final state allows a peculiar
version of quantum ``teleportation" \cite{kn:bais} to salvage the
information that is apparently lost in a black hole. The proposal
amounts to a simple concrete expression of the idea that no
information must be allowed to leave the interior spacetime through
the singularity
---$\,$ or through whatever replaces the singularity in a more
complete theory\footnote{Henceforth, in order to avoid repeating
this long locution, we shall use ``spacetime edge" to mean any of
the spacelike regions which are singular in classical general
relativity but where some other description takes over in a more
complete theory. In a Euclidean-gravity description, for example,
such edges might be the ones along which the Lorentzian/Euclidean
transition occurs. For a dissenting view on the transmissibility of
information through black holes, see \cite{kn:smolin}.}.

The Horowitz-Maldacena proposal is a very natural and elegant
approach to the black hole information problem. Furthermore, it has
the great virtue of locating the new physics near to the spacetime
edge; so it does not ask us to believe that classical general
relativity needs a drastic revision at energy scales where it is
extremely well-tested. It therefore came as a surprise when a
serious objection was raised against it \cite{kn:preskill}.
Gottesman and Preskill argue that, just inside the event horizon
---$\,$ that is, long before the spacetime edge is approached
---$\,$ the collapsing star which forms the black hole will become
entangled with the infalling Hawking radiation. This leads to a
non-trivial interaction between \emph{past and future} versions of
the relevant information, and this in itself can lead to violations
of unitarity. One might hope that a \emph{small} modification of the
Horowitz-Maldacena condition at the spacetime edge can compensate
for this, restoring the unitarity of the black hole S-matrix. But
Gottesman and Preskill argue that this would require an
``\emph{implausible conspiracy}" \cite{kn:preskill} between
conditions near to the event horizon and the much \emph{later} state
near to the spacetime edge.

The reliance of the Horowitz-Maldacena proposal on an ``implausible
conspiracy" between conditions at different times and energy scales
is very much reminiscent of the problem of understanding \emph{the
arrow of time}, a problem which has recently attracted renewed
attention\footnote{See
\cite{kn:dyson}\cite{kn:albrecht}\cite{kn:carroll}\cite{kn:wald}\cite{kn:steve}\cite{kn:banks}
for general background, \cite{kn:arrow}\cite{kn:baby}\cite{kn:BBB}
for the point of view advocated here, and
\cite{kn:bojowald}\cite{kn:mathur} for recent alternative
perspectives.}. As we trace the history of the Universe back towards
its beginning, we witness an ``implausible conspiracy" on a
cosmological scale as the entropy ---$\,$ particularly the
gravitational contribution to it \cite{kn:penrose}
---$\,$ steadily declines, ultimately to fantastically small values.
Thus, we \emph{know} that apparently wildly improbable correlations,
leading to an extraordinarily non-generic state at a far-off time
and a much higher energy scale, do occur in our Universe as this
spacetime edge is approached. The question is whether a [perhaps
much milder] version of this can occur inside a black hole. If it
can, this might shed new light on the status of the
Horowitz-Maldacena proposal.

Horowitz and Maldacena were well aware that there is a fundamental
relationship between the information loss problem and the arrow of
time\footnote{A very detailed study of this link, taking a different
point of view from the one advocated here, has been given in
Reference \cite{kn:festuccia}.}. Similarly, Gottesman and Preskill
point out that one way of understanding the Horowitz-Maldacena
proposal is in terms of information propagating \emph{backwards in
time} from the spacetime edge \cite{kn:preskill}. Nevertheless
Horowitz and Maldacena argue that the arrow does \emph{not}
completely reverse inside a black hole. Indeed, a complete
reversal\footnote{By a ``reversal" of the arrow, we really just mean
that the arrow points in an unexpected direction, that is,
\emph{away} from the black hole spacetime edge. This makes sense
whether or not there is an arrow in the external spacetime.} of the
arrow inside a realistic black hole, embedded in an external
spacetime which itself has an arrow, could lead to serious
difficulties, since one might have to impose both initial and final
conditions on \emph{every} system inside the black hole; and it is
far from clear that this can always be done in a way consistent with
a quantum-mechanical coarse-graining of phase space \cite{kn:zeh}.

In the present case, however, we do not need a \emph{complete}
reversal of the arrow inside the black hole, since the effect of the
entanglement pointed out in \cite{kn:preskill} is extremely small
compared to the overall entropy of the hole. In fact, there is no
reason to think that reversing the arrow of time inside the black
hole has to be an ``all-or-nothing" matter; all we need here is that
there should be \emph{some} traces of such an effect.

It is now widely accepted that the behaviour of \emph{Anti-de
Sitter} black holes will provide the key to an explicit
string-theoretic resolution of the black hole information problem.
The duality of the thermodynamics of \emph{these} black holes with
thermal aspects of a conformal field theory does strongly suggest
that their evolution is completely unitary [particularly in the case
of eternal black holes \cite{kn:juan}]; and of course there is ample
evidence \cite{kn:son} that these objects really do have some deep
physical significance. Despite all this, the precise way in which
properties of AdS black holes constrain the behaviour of black holes
in the real world, where the cosmological constant is positive, is
not yet fully understood. To judge by the methods used in
\cite{kn:juan}, it seems likely that this has to be done by passing
through the Euclidean domain. But since Euclidean methods invariably
involve a ``sum over geometries"
---$\,$ and this sum is in fact \emph{necessary} for the maintenance
of unitarity \cite{kn:andy}\cite{kn:hawk}
---$\,$ this probably means that we must not expect a direct
one-to-one correspondence between AdS black holes and those in de
Sitter spacetime. Instead, we should expect to use some kind of
``sum" over the Euclidean versions of \emph{all} AdS black holes of
a given entropy to probe the de Sitter black hole with that value
for the entropy. As we shall explain in detail, the entropy is
\emph{far} from being able to specify a unique black hole in AdS, so
the correspondence is indeed not one-to-one. [Note that the
information loss problem is as serious for eternal black holes as it
is for any other variety, since such holes can apparently convert
pure states to thermal ones; so eternal black holes must be included
in the ``sum".]

With this in mind, it is interesting to ask: are there \emph{any}
AdS black holes inside which the arrow \emph{does} reverse? If there
are, these objects might make a small contribution to the sum over
AdS black holes with a given entropy, and this could point the way
towards arranging a minor ``conspiracy" of the kind described by
Gottesman and Preskill. The hope is that this might allow the
Horowitz-Maldacena proposal to work.

We begin by discussing the range of possible AdS$_5$ black hole
geometries. [For the sake of definiteness, and because this is the
case that is best understood, we focus on five-dimensional
spacetimes, but this is not essential.] As is well known, this range
is much wider than in the case of asymptotically flat or de Sitter
spacetimes: one has black holes with event horizons which are
\emph{not positively curved}. These ``topological" black holes
\cite{kn:mann}\cite{kn:danny} are just as valid as the more familiar
spherical ones at the \emph{perturbative} level in string theory,
and we clearly need to know whether all of them are present in the
full theory. Fortunately, in string theory one has a simple yet
powerful technique which eliminates a large subset of these
``topological" black holes: we just have to apply the
[non-perturbative] brane pair-production stability criterion of
Seiberg and Witten \cite{kn:seiberg}; see \cite{kn:maoz} and
\cite{kn:porrati} for clear introductions to and applications of
this remarkable method. Using this method, we find in Section 2 that
a black hole with a negatively curved event horizon is always
unstable [in the Seiberg-Witten sense] in string theory. This
eliminates one entire class of topological black holes from
consideration.

The survivors are the AdS black holes having a torus, or a
non-singular quotient of a torus, as their event horizon. In section
3 we study these toral black holes and show not only that they are
stable, but that they remain stable as long as the Null Energy
Condition holds. This condition is expected to be valid here,
because toral black holes always have a positive specific heat and
are able to come into equilibrium with their own Hawking radiation:
they are \emph{eternal}, like the black holes studied in
\cite{kn:juan}. Thus, we have to take toral black holes as seriously
as their spherical counterparts.

In section 4, we briefly review the theory of the arrow of time
advanced in \cite{kn:arrow}\cite{kn:baby}\cite{kn:BBB}, and ask
whether it implies that the arrow might ``reverse" inside an AdS
black hole. The answer is that it does indeed imply this for
\emph{toral} black holes, but \emph{not} for spherical ones. The
argument makes use of the deep mathematical work of Gromov and
Lawson on ``weakly enlargeable" manifolds [see \cite{kn:lawson}.] If
a correspondence between realistic black holes and some kind of sum
over their AdS counterparts of the same entropy does exist, then one
may ultimately be able to use toral AdS black holes to arrange a
``conspiracy" of the kind needed to ensure the unitarity of the
final state S-matrix.

\addtocounter{section}{1}
\section* {\large{\textsf{2. No Negatively Curved Event Horizons in String Theory}}}
Consider an asymptotically AdS$_5$ black hole, with a fixed
entropy\footnote{Our definition of ``asymptotically AdS black hole"
requires only [in the vacuum case] that infinity be timelike and
that the Euclidean version be conformally compactifiable; this is
motivated by the AdS/CFT correspondence. We shall not consider
``black rings" \cite{kn:empreall} here. We shall throughout apply
the words ``area" and ``volume" somewhat inconsistently, with a view
to intuition rather than precision of language.}. In AdS$_5$, the
specification of the entropy does not fix a unique spacetime metric,
because the field equations do not enforce a particular foliation of
the spatial hypersurfaces. Suppose that these hypersurfaces outside
the event horizon are foliated by sections modelled on some specific
compact three-dimensional space, C$_k$, of constant curvature k,
where k = $\{- 1, 0, + 1 \}$. Then all of the following
metrics\footnote{Our emphasis on constant curvature is based on the
following fact: if we impose the reasonable condition that the local
geometry should become indistinguishable from that of AdS$_5$ when
the black hole mass tends to zero, then \cite{kn:danny} the event
horizon \emph{must} be a space of constant curvature.} are
\emph{equally valid} as solutions of the field equations [with no
matter other than the [negative] vacuum energy] \cite{kn:danny}:
\begin{equation}\label{A}
\m{g(BH_k) = - \Bigg[{r^2\over L^2}\;+\;k\;-\;{16\pi M\over
3A_kr^2}\Bigg]dt^2\;+\;{dr^2\over {r^2\over L^2}\;+\;k\;-\;{16\pi
M\over 3A_kr^2}} \;+\; r^2d\Omega_k^2}.
\end{equation}
Here L is the radius of curvature of AdS$_5$, and $\m{d\Omega_k^2}$
is a metric of constant curvature k on the three-dimensional space
C$_{\m{k}}$; then A$_{\m{k}}$ is the area of this space.

Before we proceed, let us consider a reasonably realistic black
hole, which will be asymptotically de Sitter, not asymptotically
AdS. Idealised versions of these objects have singularities in the
past as well as the future, but the past singularities do not occur
in realistic versions. Instead, a real black hole is formed from the
collapse of a star; this eliminates the past singularity. In order
to encode this important property of real black holes, we shall take
an orbifold quotient of each of the spacetimes with metrics given in
equation (\ref{A}), factoring out the obvious ``horizontal fold"
isometry which maps the past singularity onto the future
singularity; in effect, t runs from zero to infinity outside the
event horizon [though in a classical hole it still runs from $-
\infty$ to $+ \infty$ inside, where it is a spatial coordinate]. In
each case, then, there is only \emph{one} spacetime edge to
consider. The conformal diagrams are then similar to the upper half
of the AdS black hole conformal diagram discussed in
\cite{kn:veronika}.

Returning to equation (\ref{A}): it is important to understand that
A$_{\m{k}}$ is not yet uniquely defined, because for each k there
are \emph{many} spaces of constant curvature k. In the case of k =
$1$, A$_1$ is fixed by the \emph{topology} of C$_1$; for example if
C$_1$ is the unit radius three-sphere, it is equal to 2$\pi^2$,
while for the unit radius real projective space $\bbr P^3$ it would
be $\pi^2$, and so on for all of the other [infinitely numerous]
three-dimensional manifolds of unit positive constant curvature
\cite{kn:wolf}. We shall return to this point in Section 5.

In the case where k = 0, there is again a topological ambiguity
[there are six possible topologies in the orientable case
\cite{kn:conway}], but, in addition, the size and shape of the
compact flat space can be freely prescribed. The simplest
possibility is to declare that a given value of the coordinate r
corresponds to a \emph{flat cubic torus}, defined as the Riemannian
product of three circles each of radius Kr, where K is a positive
dimensionless number. In this special case, the area of the surface
r = constant at a given time outside the black hole is
8$\pi^3$K$^3$r$^3$, so A$_0$ = 8$\pi^3$K$^3$. We shall take this as
the definition of K in general, that is, for arbitrary compact flat
three-dimensional manifolds. K is then a measure of the overall
relative size of the space. The presence of such a continuous
parameter distinguishes the flat case in a fundamental way from the
other two categories.

The case of k = $- 1$ is the most difficult and potentially
troublesome one, because there is a vast set of distinct compact
manifolds of unit negative curvature [see \cite{kn:weeks}]. Fixing
the entropy of the black hole now requires having a complete
knowledge of the spectrum of possible volumes of these manifolds. If
such black holes\footnote{The perturbative behaviour of these black
holes has been studied in depth; see \cite{kn:siopsis} and its
references.} have to be included in the sum over AdS black holes of
given entropy, then we will have to confront a formidable
mathematical problem; also, these objects might dominate the sum,
which would be hard to understand. Fortunately, we shall see that
this is not the case if we work within the constraints imposed by
string theory.

All of the metrics in equation (\ref{A}) have Euclidean versions, in
which the complexified version of t has to parametrise a circle;
thus the Euclidean metric, g(EBH$_{\m{k}}$), is a metric on a
manifold with the topology of S$^1\,\times \bbr \,\times \m{
C_{\,k}}$. At large distances, the circumference of the circle is a
constant multiple of r; the dimensionless constant of
proportionality, P, must be chosen so that the Euclidean metric is
not singular at r$_{\m{eh}}$, the value of r at the event horizon.

Now we wish to consider these spaces in the string context, which
means that we have to consider the possible nucleation of branes. In
the case of a three-brane in any Euclidean asymptotically AdS$_5$
space, we can write the action in the form \cite{kn:seiberg}
\begin{equation}\label{B}
\m{S\;=\;\Theta\,\Big\{Brane \,Area\Big\}\;-\;\mu\Big\{Volume
\,Enclosed \, by \, Brane\Big\},}
\end{equation}
where $\Theta$ is the tension and $\mu$ is a constant related to the
charge. Non-perturbative instabilities arise
\cite{kn:maoz}\cite{kn:porrati} if the action becomes negative; the
BPS case is the most dangerous one. In that case, $\mu$ is just
4$\Theta$/L, where L is the curvature radius of the asymptotic
AdS$_5$, so we can compute the action if we know the volume and area
of a brane located at some value of r. In the case of the Euclidean
version of the metric\footnote{In this metric, the coefficient of
dt$^2$ is the reciprocal of the coefficient of dr$^2$; the resulting
cancellation explains why the volume integral is so easily
evaluated.} given in equation (\ref{B}), we have, for a brane
located at coordinate value r,
\begin{equation}\label{C}
\m{S(r ; M,L,\Theta, K (k = 0)) \;=\;\Theta P L A_k\Bigg\{
r^3\Bigg[\,{r^2\over L^2}\;+\;k\;-\;{16\pi M\over
3A_kr^2}\,\Bigg]^{1/2}\;-\;{r^4\,-\,r_{eh}^4\over L}\Bigg\}.}
\end{equation}
Here the notation means that the action is determined by r, M, L,
$\Theta$, k, and, if k = 0, also by K. [r$_{\m{eh}}$ and P are
determined by M, L, and, where applicable, K.]
\begin{figure}[!h] \centering
\includegraphics[width=0.7\textwidth]{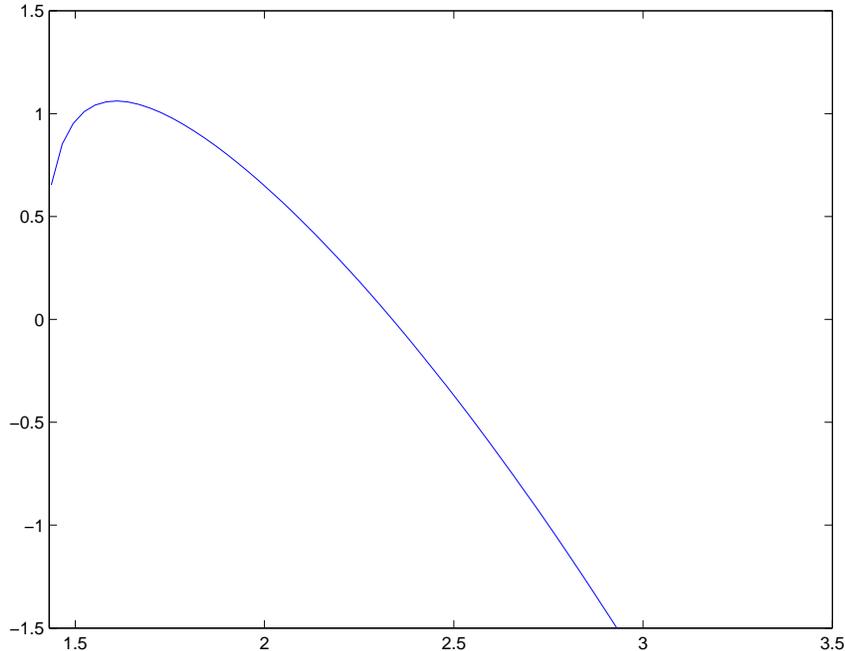}
\caption{Brane Action, Negatively Curved Event Horizon.}
\end{figure}
The simple form of equation (\ref{C}) allows us to write the action
as
\begin{equation}\label{D}
\m{S(r ; M,L,\Theta, K (k = 0)) \;=\;\Theta P L
A_k\Bigg\{{L[kr^2\;-\;{16\pi M\over 3A_k}]\over
1\;+\;\Bigg[1\;+\;{kL^2 \over r^2}\;-\;{16\pi ML^2\over
3A_kr^4}\Bigg]^{1/2}}\;+\;{r_{eh}^4\over L}\Bigg\}}.
\end{equation}

When k = 1, it is easy to show that this expression is positive for
all values of r larger than r$_{\m{eh}}$, and in fact it diverges to
$+\,\infty$ for large r. Thus the familiar spherical AdS black
holes, and all of their less familiar relatives with event horizons
having the topology of some non-singular quotient of the
three-sphere, are completely stable against this effect, and all of
them must be retained in string theory. We shall return to them
later.

For k = $-1$, the action is again positive at first. Soon, however,
the graph [Figure 1, with representative parameter values] turns
over, and in fact the action is unbounded below. Thus, \emph{black
holes with negatively curved event horizons are non-perturbatively
unstable in string theory}; that is, they are not solutions of the
full theory. The power of the Seiberg-Witten technique is underlined
by the fact that a simple calculation entails such a strong
conclusion.

In fact, Seiberg and Witten showed that this kind of instability
will \emph{always} arise if the scalar curvature at
infinity\footnote{That is, the scalar curvature of a metric
representing the conformal structure of the boundary; this scalar
curvature can be taken to be constant, without loss of generality
\cite{kn:schoenyam}.} is negative. In the case at hand, the
conformal structure at infinity is represented by the metric
\begin{equation}\label{DD}
\m{g(EBH_{\,- 1}, r \rightarrow \infty) \;=\; dt^2 \;+\;
L^2d\Omega_{\,- 1}^2},
\end{equation}
which is obviously a metric of constant negative scalar curvature on
a compact manifold of topology S$\m{^1\,\times C_{\,- 1}}$. The
basic point is that a negative scalar curvature at infinity amounts
to having a negative (mass)$^2$ for certain scalar fields there;
since the boundary theory is therefore unstable, the dual theory in
the bulk must likewise be unstable.

We see that the requirement of non-perturbative stability gives an
enormous reduction in the number of AdS black holes which need to be
considered in a string-theoretic approach. But one class of
``topological" black holes remains to be considered: those with k =
0. Notice that the Seiberg-Witten duality argument does not work
here, since the scalar field at infinity simply becomes massless. In
fact, asymptotically AdS spaces with Euclidean versions having
compact flat boundaries can be either stable or unstable against
brane nucleation, depending on the details. [Explicit examples
displaying instability were first presented in \cite{kn:unstable}.]
We now proceed to discuss these black holes.

\addtocounter{section}{1}
\section* {\large{\textsf{3. Black Holes with Flat Event Horizons}}}
\begin{figure}[!h] \centering
\includegraphics[width=0.7\textwidth]{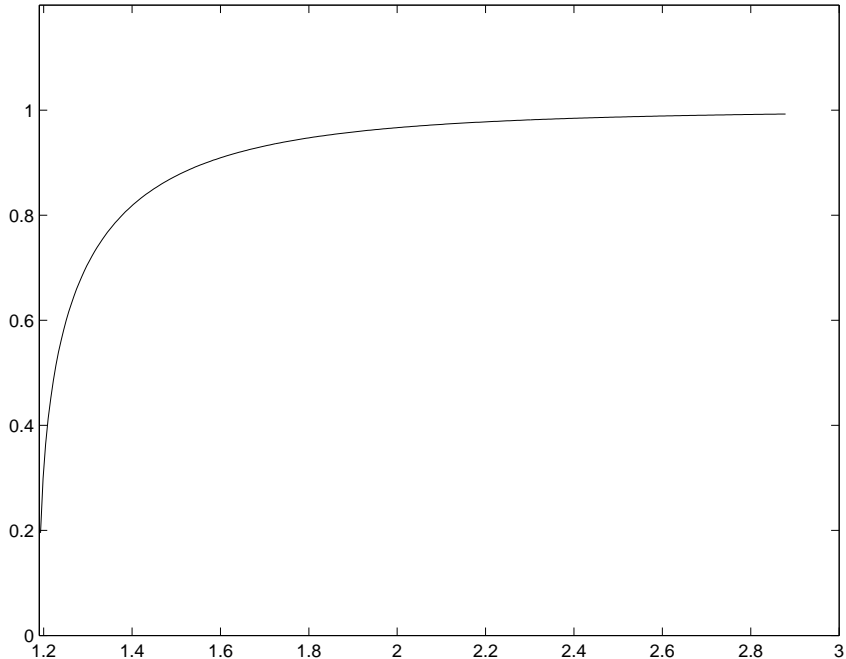}
\caption{Brane Action, Flat Event Horizon.}
\end{figure}
The graph of the brane action in this case has an interesting form:
see Figure 2. The action increases from zero, but it does not
diverge either positively or negatively: it is asymptotic to a fixed
constant \emph{positive} value. Thus, these black holes are stable
against brane pair-production, unlike their counterparts with
negatively curved event horizons.

The value of r at the event horizon, r$_{\m{eh}}$, is given by
\begin{equation}\label{E}
\m{r_{eh}\;=\;\Bigg[{16\pi ML^2\over 3A_0}
\Bigg]^{1/4}\;=\;\Bigg[{2\, ML^2\over 3\pi^2K^3} \Bigg]^{1/4}.}
\end{equation}
Thus r$_{\m{eh}}$ actually decreases as K increases. However, the
area of the event horizon is
\begin{equation}\label{F}
\m{A_{eh}\;=8\times [2/3]^{3/4}\times \pi^{3/2}\times
\big[MK\big]^{3/4}L^{3/2}\;\;\approx\;32.866\times
\big[MK\big]^{3/4}L^{3/2},}
\end{equation}
which does increase with K. Notice that the size of the event
horizon is \emph{not} fixed by the mass and the AdS radius of
curvature, as it is in the case of a spherical horizon; the
relevant parameter here is MK.

It is possible to show \cite{kn:surya} [see also
\cite{kn:gall1}\cite{kn:gall2}] that the entropy of these black
holes is, as usual, proportional to the area of the event horizon;
thus one way to think about K is as a measure of the black hole
entropy for fixed mass. More remarkably, the specific heat of these
black holes is \emph{also} a constant multiple of the area of the
event horizon \cite{kn:surya} [and is independent of the
temperature], so K might alternatively be regarded as a measure of
the specific heat for fixed mass.

It is clear from this discussion that the specific heat of a toral
black hole is always positive, for \emph{all} parameter values. Such
black holes do not evaporate away completely: they come into
equilibrium with their own Hawking radiation, just as ``large" black
holes do in the spherical case \cite{kn:wittenconf}. This means that
they are \emph{eternal}.

Eternal black holes are of great interest in the context of the
information loss problem, because they dominate the relevant
canonical thermodynamic ensemble in the spherical case, and, perhaps
more importantly, because they are the ones which definitely have a
specific CFT dual. They are therefore the ones for which we have the
strongest evidence regarding the preservation of unitarity. For
example, Witten \cite{kn:wittenconf} stresses that the specific heat
of the CFT is positive, so the AdS/CFT correspondence is most direct
for eternal black holes. Note that a small object in a pure state
which is thrown into an eternal black hole will have its mass
radiated in a thermal state according to the usual calculation; thus
the information loss problem is just as serious for eternal black
holes as for any other variety. The precise form taken by the
information loss problem in the case of eternal black holes was
discussed in Maldacena's classic study \cite{kn:juan}.

Later we shall see that eternal black holes are in fact the most
important ones for the purposes we have in view in this work. Notice
in this connection that, for a \emph{small} realistic black hole,
there may be \emph{no} eternal AdS black hole of that entropy in the
spherical case; however, there will always be an eternal toral AdS
black hole for any value of the entropy\footnote{Here, for
simplicity, we are ignoring the effects of phase transitions in the
dual CFT: see \cite{kn:surya}.}.

The asymptotic value of the action for flat branes is
\begin{equation}\label{G}
\m{S(\infty ; M,L,\Theta, K) \;=\;{8\over 3}\pi \Theta P M L^2,}
\end{equation}
and it can be shown \cite{kn:danny} that the value of P [giving the
size of the circle defined by Euclidean ``time"] is
\begin{equation}\label{H}
\m{P\;=\;{\pi L\over r_{eh}}}
\end{equation}
in this case. Thus we have, from equation (\ref{E}),
\begin{equation}\label{I}
\m{S(\infty ; M,L,\Theta, K)\;=\;{8\over 3}\times [3/2]^{1/4}\times
\pi^{5/2}\times \Theta L^{5/2} [MK]^{3/4} \;\approx\;51.626\times
\Theta L^{5/2} [MK]^{3/4}.}
\end{equation}
Notice that
\begin{equation}\label{J}
\m{{S(\infty ; M,L,\Theta, K)\over \Theta} \;=\;{\pi L A_{eh}\over
2},}
\end{equation}
so the asymptotic brane action per unit tension is essentially just
the entropy or the specific heat of the black hole. That the branes
at large distances contain information about the entropy of the hole
might be regarded as an example of holography. We take it as further
evidence that unitarity is maintained here, since the [unitary] CFT
on the boundary should encode all data carried by branes propagating
towards infinity.

While it is clear that \emph{vacuum} black holes with flat compact
event horizons are stable against the nucleation of branes, it is
also clear that this stability is not as secure as in the positively
curved case. In the latter, the action increases rapidly and
diverges towards $+\infty$, so while the introduction of matter
[including the effects of Hawking radiation] into the spacetime will
change the details, it is unlikely to change the qualitative
behaviour, particularly for large values of r. In the case of flat
event horizons, however, the action is always finite, and it is
always small if the entropy of the black hole is small. Experience
in the cosmological case \cite{kn:unstable}\cite{kn:baby} shows
that, in such cases, there can be a danger that the action will
become negative at large r. On the other hand, in \emph{all} cases
---$\,$ including black holes with positively curved event horizons ---$\,$
the action is close to zero near to r = r$_{\m{eh}}$. One might be
concerned that, if the spacetime around the black hole is not
exactly empty, then the action might become negative either at very
large values of r or near to the event horizon.

In some cases one can see directly that this will not happen. For
example, it does not happen if the black hole carries a reasonable
amount of electric charge. Here, ``reasonable" means ``small
compared to the value which causes the black hole to become
extremal," that is, the value beyond which Cosmic Censorship is
violated. In fact, it is possible to show that an AdS black hole
with a flat event horizon remains completely stable against brane
nucleation when charge is added to it, up to the point where the
relevant dimensionless parameter is equal to about 92$\%$ of its
value in the extremal case. The detailed structure of AdS
Reissner-Nordstr$\ddot{\m{o}}$m black holes with flat event horizons
is of some independent interest, and will be discussed elsewhere.

Again, experience in the cosmological case
\cite{kn:unstable}\cite{kn:baby} leads us to expect that problems
with negative brane actions are often associated with violations of
energy conditions. We can investigate this possibility in a simple
way by studying a ``toral star", that is, a perfect fluid in a
spacetime having a similar structure to that of the vacuum toral
black hole. The metric has the form
\begin{equation}\label{K}
\m{g(TS) = - \,f(r,M,L,K)dt^2\;+\;h(r,M,L,K)dr^2 \;+\;
r^2d\Omega_0^2},
\end{equation}
where f and h are functions to be determined, and we confine
ourselves to values of r greater than or equal to r$_{\m{eh}}$ so as
to be able to make a comparison with the black hole. [We can imagine
that the ``star" is about to collapse into the black hole, so r =
r$_{\m{eh}}$ represents a surface inside the ``star".] We assume
that the fluid is distributed so that f and h approach their vacuum
toral black hole values at large r.

The brane action in this case is
\begin{equation}\label{L}
\m{S(r ; M,L,\Theta, K) \;=\;8 \pi^3 \Theta P L K^3\Big\{
r^3\,f^{1/2}\;-\;{4\over L}\int_{r_{eh}}^r \rho^3\,[\,f\,h\,]^{1/2}
d\rho\Big\}.}
\end{equation}
Since the ``star" is not a black hole, the function f is strictly
positive at r$_{\m{eh}}$ and near to it, so the first term on the
right side has the effect of increasing the action away from the
value it would have for a vacuum black hole. The dangerous term is
the second one: the function fh is identically equal to unity for a
black hole, but not for a fluid \cite{kn:jacob}.

Now the radial null vectors in this geometry, which we denote by
n$^{\mu}$, can be taken to have the form
\begin{equation}\label{M}
\m{n^{\mu}\;=\;(h^{1/2},\,\pm\,f^{1/2},\,0,\,0,\,0)^T,}
\end{equation}
and one can show \cite{kn:jacob} that, if R$_{\mu\nu}$ is the
Ricci tensor,
\begin{equation}\label{N}
\m{R_{\mu\nu}\,n^{\mu}\,n^{\nu}\;=\;{3\,[\,f\,h\,]'\over 2\,r\,h},}
\end{equation}
where the dash denotes a derivative with respect to r. But the
\emph{Null Ricci Condition} or NRC is the statement that the Ricci
tensor satisfies
\begin{equation}\label{O}
\m{R_{\mu\nu}\,n^{\mu}\,n^{\nu}\;\geq\;0}
\end{equation}
for all null vectors. If the Einstein equations hold, the NRC is
equivalent to the Null Energy Condition or NEC, the weakest of the
classical energy conditions, which just requires the
stress-energy-momentum tensor T$_{\mu\nu}$ to satisfy
$\m{T_{\mu\nu}\,n^{\mu}\,n^{\nu}\;\geq\;0}$ for all null vectors.
Assuming that these conditions hold, we see that, if fh is not
identically equal to unity, it has always to increase towards its
asymptotic value, which is unity; therefore it must always be
smaller than its value in the vacuum case, which is identically
equal to unity. Since the domain of integration is the same and the
integrand becomes smaller if matter satisfying the NEC is
considered, the effect of the fluid is to diminish the second term
on the right side of equation (\ref{L}).

We conclude that the right side of (\ref{L}) increases from its
black hole value if the fluid satisfies the NEC, and so the danger
of instability due to uncontrolled pair-production of branes is
reduced, never increased, under this assumption. The same conclusion
holds for black holes with positively curved event horizons.

The reader might object at this point that while all may be well at
the classical level, Hawking radiation, with which we are of course
directly concerned in this work, is often associated precisely with
\emph{violations} of the NEC. This actually does happen in the most
familiar case, in which the black hole shrinks as it radiates. For,
in that case, photons which are inside the apparent horizon at one
point will find themselves outside it subsequently, leading to an
expansion of the corresponding null congruence; this violates the
NEC, by the null version of the Raychaudhuri equation
\cite{kn:strominger}.

Toral black holes, however, do not shrink to arbitrarily small size:
their specific heat, being a positive multiple of the area of the
event horizon, is always positive for all mass values. As we
discussed earlier, they settle down to a static equilibrium with
their own Hawking radiation. Thus we do not need to be concerned
about NEC violation for ``large" black holes in the case of
positively curved event horizons, or for any black hole in the flat
case. On the other hand, there will be strong violations of the NEC
in the final stages of the evaporation of a ``small" spherical AdS
black hole, and the brane action could well become negative in that
situation.  By its very nature, however, this kind of black hole is
a transient phenomenon; when the evaporation is complete, the NEC is
no longer violated, and the brane action will cease to be negative.
Thus we cannot conclude that ``small" spherical AdS black holes
should be excluded from the sum over black holes\footnote{AdS black
holes with negatively curved event horizons are, by contrast,
eternal, so the brane action is permanently negative in that case.}.
However, eternal black holes are clearly easier to understand, so we
shall focus on them for the remainder of this work.

To summarize: ``large" AdS black holes with positively curved
event horizons, and all AdS black holes with flat event horizons,
settle down to a static state which is non-perturbatively stable
in string theory. In particular, we conclude that if we need to
survey all AdS black holes which are compatible with string
theory, we have to include the flat case. As we shall now argue,
the geometric and topological differences between AdS black holes
with the two kinds of possible event horizons lead to drastically
different thermodynamic behaviour in their interiors.

\addtocounter{section}{1}
\section* {\large{\textsf{4. Time Inside AdS Black Holes}}}

We now begin our search for traces of the kind of ``conspiracy"
which Gottesman and Preskill demand if the Horowitz-Maldacena
proposal is to be made to work. This raises fundamental issues
regarding the nature of time inside AdS black holes.

The question as to the \emph{direction of time} inside black holes
has been debated extensively \cite{kn:price}. The point is this.
There are strong arguments \cite{kn:BBB} to the effect that the
observed arrow of time arises from some property of the spacetime
edge corresponding to the creation of the Universe: the entropy
associated with this edge was extraordinarily low, and the arrow is
due to the natural evolution towards more generic states. One
naturally asks: why should the arrow not point away from \emph{all}
spacetime edges, leading to a ``reversal" of the arrow of time
inside black holes? Conversely, if one has an argument which
``proves" that the arrow cannot point away from the spacetime edge
inside a black hole, then one must immediately explain why the same
argument does not ``prove" that our Universe should have no arrow at
all
---$\,$ that is, why the argument cannot be applied to the spacetime edge at
which the Universe was created\footnote{If the ultra-low-entropy
conditions at the beginning of our Universe were set up by some
still earlier state \cite{kn:carroll}\cite{kn:baby}, then one either
applies the same argument to that state, or one has to explain why
black hole singularities are resolved in a radically different way
to their apparently similar cosmological counterparts.}.

We see from this that all such questions \emph{can only be answered
in the context of a specific theory of the arrow of time}. In such a
theory one might find that an arrow pointing away from the spacetime
edge exists for ``initial" cosmological edges but not for any black
hole edge [as Penrose postulates \cite{kn:penrose}], or that it
occurs in some black holes contributing to a ``sum over geometries"
but not in a way that dominates the sum. The point is that such
conclusions must be a matter of deduction from a specific theory;
the question cannot be settled by appealing to standard
statistical-mechanical expectations. For we \emph{know} that
``standard statistical-mechanical expectations" [to the effect that
high-entropy states are generic] are \emph{not} realised in the case
of the only spacetime edge to which we have some observational
access
---$\,$ the one associated with the Big Bang. It is therefore hard
to see why they should be realised inside black holes.

One of the main observations of the present work is that the
Gottesman-Preskill argument is based on the \emph{assumption} that
the arrow of time can never point away from a spacetime edge inside
a black hole. They state, for example, that ``the interior of a
black hole is a tumultuous place". This is indeed the case under
``normal" circumstances. Gravitational entropy is not fully
understood, but it is clearly associated with the degree to which
spatial sections are anisotropic and inhomogeneous. \emph{If} the
arrow of time inside a black hole points \emph{towards} the
spacetime edge, then one will find that a small perturbation of the
geometry inside the hole will cause the internal spatial sections to
become more and more inhomogeneous and anisotropic as the classical
``singularity" is approached, a tendency which becomes still more
conspicuous as the spatial sections contract [because this increases
the entropy density]. This is precisely the behaviour computed from
classical general relativity \cite{kn:imponente}, and it is the
picture that Gottesman and Preskill have in mind. But \emph{if} the
arrow points \emph{away} from the black hole spacetime edge, then a
small perturbation will become \emph{smaller} as the edge is
approached, by means of what would look like an ``implausible
conspiracy".

In the case of a \emph{realistic} black hole, embedded in an
asymptotically de Sitter spacetime which itself has an arrow of
time, the assumption made by Gottesman and Preskill is entirely
reasonable. For if we try to maintain two independent arrows
pointing in opposite directions, producing two different but
consistent accounts of the evolution of any given system which
enters the black hole, we will find that this requires fine-tuning
on a scale below the Planck volume in phase space \cite{kn:zeh}. In
reality, one would expect the system with the larger number of
degrees of freedom ---$\,$ the outside world
---$\,$ simply to overwhelm the system with fewer; at most one would
find inside the black hole some \emph{traces} of a reversal. But
that is all we need. The problem is to find a way of describing
these ``traces"; the hope is that AdS black holes may provide this
description.

Consider first \emph{de Sitter} spacetime. This spacetime is said to
be \emph{globally hyperbolic} [\cite{kn:waldbook}, Chapter 8], which
essentially means that it has spacelike surfaces [Cauchy surfaces]
on which the prescription of initial data determines the evolution
of matter and geometry for all subsequent time. Global hyperbolicity
is clearly essential for the existence of an arrow of time of the
kind we observe, which is apparently ubiquitous in both space and
time. Such an arrow arises only because the prescription of
ultra-low-entropy conditions on \emph{one} Cauchy surface sufficed
to enforce the second law of thermodynamics for all time. In fact,
the existence of a universal arrow of time is the strongest evidence
we have for the global hyperbolicity of our Universe.

When we turn to \emph{AdS} black holes, we find that the environment
is very different. Asymptotically AdS spacetimes are \emph{not}
globally hyperbolic; the future is not determined by data on a
single Cauchy surface, since information can enter from infinity. It
follows that such spacetimes do not, in general, have a
\emph{universal} arrow of time of the kind we observe in our
Universe. This is reflected in the fact that asymptotically AdS
spacetimes have a timelike Killing vector field defined everywhere
except perhaps near and inside black holes.

An asymptotically AdS spacetime, left to its own devices, will
therefore not contain stars; for stars are very low-entropy
systems which, in our Universe, inherit that property from the
systems of still lower entropy which characterized the Big Bang:
they owe their existence to the ubiquity of the arrow of time.
Thus we should not think of AdS black holes as forming in the same
way as realistic black holes. We should instead think of them as
systems which we ``prepare" or ``set up" in a background spacetime
\emph{with no global arrow of time}. Note that, in string theory,
this ``setting up" cannot be done in an arbitrary way; it has to
be performed under the strict constraints imposed by the AdS/CFT
correspondence \cite{kn:veronika}\cite{kn:hong}. This keeps the
data entering from infinity under control. [For reasons explained
when we first discussed equation (\ref{A}), we do this ``setting
up" in such a way that there is only \emph{one} spacetime edge.]

The situation is particularly clear in the case of \emph{eternal}
AdS black holes: the hole is in equilibrium with a static gas of
Hawking radiation, so it is natural to regard the exterior spacetime
as having no arrow. \emph{Inside} the black hole, however, there is
no timelike Killing field, and so it becomes possible to imagine
that a \emph{local} arrow could exist there, pointing either towards
or away from the spacetime edge\footnote{A complete AdS/CFT
description of the geometrical dynamics inside a black hole is not
available; see however \cite{kn:hertog} for a discussion of one
possible way in which an arrow of time might arise in singular
asymptotically AdS spacetimes.}. If this local arrow points away
from the spacetime edge, then it is possible to think of the
resulting object as a ``white hole"; but, as is well known
\cite{kn:whitehawk}, there is little to be gained from such a
description in a quantum-mechanical treatment, and those
observations are reinforced in this case by AdS/CFT considerations.
We shall therefore not use this terminology.

The real point here is that if we should find a local arrow inside
an eternal AdS black hole, pointing away from the spacetime edge,
there will be no ``clash of arrows": for there is no arrow outside
the event horizon. Thus the problem of ``overdetermination"
discussed by Zeh \cite{kn:zeh} does not arise here.

The situation of \emph{non-eternal} black hole is different. Such
a hole has a local arrow defined in its vicinity just
\emph{outside} the event horizon, defined by the very process of
its complete evaporation. Note once more that this arrow arises
because of the way the hole is ``set up", as above; it is not
generated by low-entropy conditions at an initial time. An
acceptable theory of the arrow of time should predict that such
black holes do not have a ``reversed" arrow \emph{inside} the
event horizon, since this will again avoid a ``clash of arrows".

We see that the kind of behaviour we are looking for here is only to
be expected in the eternal case; also that, if it does arise in that
case, it will not lead to any paradox. If the arrow is ``reversed"
inside some kind of eternal AdS black hole, it might be possible to
show that the Horowitz-Maldacena hypothesis automatically
incorporates Gottesman-Preskill ``conspiracies".

In summary: Gottesman and Preskill make a seemingly natural
assumption about the internal thermodynamics of black holes, one
which is reasonable for de Sitter holes but perhaps not [always] in
the AdS case. We shall now investigate this question, in the
specific context of the theory advanced in
\cite{kn:arrow}\cite{kn:baby}\cite{kn:BBB}.

We begin by outlining this concrete proposal for the origin of the
arrow of time. In this approach, the original universe [which may
either be ours or one in which our universe nucleated as a
Coleman-De Luccia bubble] comes into existence along a spacetime
edge $\Sigma$ representing a transition from a Euclidean to a
Lorentzian space. This is in the manner of theories of ``creation
from nothing" \cite{kn:vilenkin}, as updated by Ooguri et al
\cite{kn:ooguri}. We assume that $\Sigma$ is a surface of minimal
volume, reflecting the idea that, in a theory exhibiting T-duality,
volumes below the value defined by the self-dual length are not
permitted.

The possible ``initial" data on $\Sigma$ are sets of objects of
the form ($\m{\rho,\,J^{a},\,K_{ab},\,h_{ab}}$), where $\rho$ is a
function, J$^{\m{a}}$ is a vector field, K$_{\m{ab}}$ is a
symmetric tensor [which is traceless in our case, because of the
minimality condition], and h$_{\m{ab}}$ is a Riemannian metric,
all defined on $\Sigma$. The number of possible distinct sets of
this kind is a measure of the initial entropy of the Universe.
Crucially, however, the number of initial data sets we need to
consider is cut down by the \emph{constraints}
[\cite{kn:waldbook}, Chapter 10], which require the sets to
satisfy
\begin{equation}\label{PP}
\m{R(h)\;=\;K_{ab}\,K^{ab}\;+\;
              16\pi\rho}
\end{equation}
and
\begin{equation}\label{PPPP}
\m{D^a\,K_{ab}\;=\; -\,8\pi J_b,}
\end{equation}
where $\m{D^a}$ is the covariant derivative operator, and R(h) is
the scalar curvature, defined by h$_{\m{ab}}$.

Now let N$^{\mu}$ be the field of inward-pointing unit normal
vectors along $\Sigma$, and let T$^{\mu}$ be the corresponding
energy-momentum flux vector. Recall that the Horowitz-Maldacena
proposal entails a requirement that no information should exit the
spacetime through the spacetime edge inside any black hole. That
appears to be a reasonable condition to impose also at the creation
of the universe; let us do so. The simplest possible mathematical
formulation of this is to demand that T$^{\mu}$ \emph{must not point
outwards from spacetime}. But when one solves the Einstein equation
with initial data as above, it turns out that K$_{\m{ab}}$ is the
extrinsic curvature of $\Sigma$, while $\rho$ is just the component
of T$^{\mu}$ parallel to N$^{\mu}$; it is in fact the \emph{total}
energy density evaluated on $\Sigma$. [Similarly, $\m{J^a}$ is the
projection onto $\Sigma$ of T$^{\mu}$.] Thus the condition that
T$^{\mu}$ must not point outwards from spacetime is just a geometric
way of formulating the condition that
\begin{equation}\label{QQ}
\rho\;\geq\;0
\end{equation}
everywhere on $\Sigma$. Note that if this is inserted into
(\ref{PP}), then every term on the right side is non-negative; so if
the left side should vanish, so must \emph{every} term on the right.

In summary: we have to consider all initial data sets of the form
($\m{\rho,\,J^{a},\,K_{ab},\,h_{ab}}$) subject to the conditions
(\ref{PP}),(\ref{PPPP}),(\ref{QQ}), and that K$_{\m{ab}}$ should
be traceless with respect to h$_{\m{ab}}$. These may seem to be
very mild constraints, and, in most cases, they are indeed mild.
Generically, then, the initial entropy of a universe created in
this way will be large, which is in accord with the usual
principles of statistical mechanics.

Remarkably, however, this conclusion is not always valid. In
particular, if [as in \cite{kn:ooguri}] the universe is created
along a surface $\Sigma$ with the topology of a \emph{torus}, then
the \emph{only} data sets satisfying the above conditions are those
of the form (0, 0, 0, p$_{\m{ab}}$), where p$_{\m{ab}}$ is a
\emph{flat} metric on the torus\footnote{The vanishing of $\rho$,
the \emph{total} energy density at the edge, means that some kind of
negative energy must be present there. In view of the toral
topology, the Casimir effect \cite{kn:coule}\cite{kn:nimah} is an
obvious candidate here.}. This follows by combining
(\ref{PP}),(\ref{PPPP}), and (\ref{QQ}) with the following extremely
deep theorem [see \cite{kn:lawson}\cite{kn:arrow}]:

\bigskip
\noindent \textsf{THEOREM (Schoen-Yau-Gromov-Lawson-Bourguignon)}:
Let h$_{\m{ab}}$ be a Riemannian metric on a torus or on any
non-singular quotient of a torus. If the scalar curvature of
h$_{\m{ab}}$ is non-negative everywhere, then h$_{\m{ab}}$ is
exactly flat.
\bigskip

We see that, if $\Sigma$ has the topology of a torus, then the
corresponding universe has an entropy on the spacetime edge which is
in fact essentially as low as it can be, at least semi-classically.
[As we discuss in the next section, flatness does not determine the
geometry of a topological torus completely, but the range of
possibilities here is obviously infinitesimal compared to the full
set of possible Riemannian metrics on the torus.] Such a universe
has an arrow of time, pointing away from the edge. This is a
realisation of [the cosmological part of] Penrose's ``Weyl Curvature
Hypothesis" \cite{kn:penrose}. The fact that this hypothesis can be
\emph{derived} as the end result of an exceedingly non-trivial
mathematical analysis is a very appealing aspect of the present
approach.

By contrast, similar techniques show that, if $\Sigma$ has the
topology of a sphere, then (\ref{PP}),(\ref{PPPP}), and (\ref{QQ})
are exceedingly weak: one way to see this is as follows. Given
\emph{any} function satisfying (\ref{QQ}), let F be \emph{any}
function on $\Sigma$ such that F $\geq$ 0 everywhere. Then it is
possible to prove [see \cite{kn:arrow} for a discussion] that the
equation
\begin{equation}\label{PPP}
\m{R(h)\;=\;F\;+\;
              16\pi\rho}
\end{equation}
\emph{always} has a solution for h$_{\m{ab}}$ if the topology of
$\Sigma$ is spherical, no matter what choices we make for $\rho$ and
F. One then merely has to solve the algebraic problem of finding all
symmetric tensors K$_{\m{ab}}$ which are traceless with respect to
this h$_{\m{ab}}$ and which satisfy F = K$_{\m{ab}}$K$^{\m{ab}}$;
obviously there are many such. Finally, if we use (\ref{PPPP}) to
define J$^{\m{a}}$, we have an initial data set which respects all
constraints. Since $\rho$ and F are completely arbitrary apart from
being non-negative, it is clear that the number of possible initial
value data sets is vast. In particular, both $\rho$ and F could be
extremely irregular, asymmetric functions, and then the metric which
solves (\ref{PPP}) will likewise be highly asymmetric, corresponding
to an arbitrarily high gravitational entropy. A universe born along
a $\Sigma$ with spherical topology will therefore \emph{not} have an
arrow.

Now let us try to apply this theory to the spacetime edge that
[presumably] replaces the singularity inside an AdS black hole. We
simply repeat the above reasoning in this case, that is, we just
consider all possible ``initial" data sets and impose
(\ref{PP})(\ref{PPPP}), and (\ref{QQ}) [and require K$_{\m{ab}}$ to
be traceless] on the edge and nothing more. [To do otherwise would
expose us to the criticism \cite{kn:price} that we are building in a
past/future distinction from the outset.] Evidently, everything
hinges on the global structure of the black hole spacetime edge.
This global structure is quite different to that of the cosmological
spacetime edge discussed above, so it is entirely possible that the
arrow of time behaves differently in this case. This is how we
answer the question raised earlier: why should the arrow not point
away from \emph{all} spacetime edges? The point is that the arrow is
intrinsically a \emph{global} phenomenon, in the geometric sense.

To determine the global structure of the spacetime edge in the AdS
black hole case, we use equation (\ref{A}) inside the event horizon,
letting r and t switch roles as usual [and remembering that t is now
allowed to take on negative values]. Whatever resolves the
singularity will of course change the metric, and we are not allowed
to assume that the spatial geometry is particularly symmetric; but
since we are only interested in the global structure at this point,
this will not matter. A spatial section near to the singularity will
be given by r = a, where a is a small constant, and the metric of
this spatial section is, from (\ref{A}),
\begin{equation}\label{R}
\m{h(r = a) = \Bigg[-{a^2\over L^2}\;-\;k\;+\;{16\pi M\over
3A_ka^2}\Bigg]dt^2 \;+\; a^2d\Omega_k^2}.
\end{equation}
Note that the coefficients of both terms are positive constants.

If we assume for the moment that t runs from $- \infty$ to $+
\infty$, then it is clear that this metric is a \emph{complete}
Riemannian metric on a manifold of topology
$\bbr\,\times\,$C$_{\m{k}}$. Here, since the metric is Riemannian,
``completeness" can be defined \cite{kn:kobayashi} either in terms
of the convergence of Cauchy sequences or in terms of
inextensibility of geodesics: it essentially just means that the
space has not been mutilated by means of arbitrary deletions.
Whether the sections are complete or not, the situation here differs
from the cosmological one considered earlier, in that the spatial
sections here are \emph{non-compact} and have infinite volume in the
complete case. [The \emph{topology} is the same in both cases; only
the geometry differs.]

As we know, ``small" black holes in the k = 1 case will evaporate
completely. In this case, the spatial sections are in fact
``mutilated" since the spatial sections inside the black hole are no
longer able to extend out towards infinity: finiteness in time
implies finiteness in space in this case. Thus the coordinate t has
a finite range here, and the spatial sections are no longer
complete. We see that the completeness of the internal spatial
sections allows us to give mathematical expression to the
distinction between eternal black holes and those which evaporate
completely\footnote{Horowitz and Maldacena \cite{kn:horomald}
suggested that the infinite volume of spatial sections in the
eternal case might prove to be important in some way.}. It does so
in a way that does not commit us to the particular, highly symmetric
metric given in (\ref{R}).

We can now formulate the question of the genericity of the black
hole edge metric as follows. We have to consider ``initial" data in
the form of sets ($\m{\rho,\,J^{a},\,K_{ab},\,h_{ab}}$), where
$\rho$, J$^{\m{a}}$, and K$_{\m{ab}}$ are defined as before, but
where now h$_{\m{ab}}$ is a metric defined on a manifold of topology
$\bbr\,\times\,$C$_{\m{k}}$, on which the Riemannian structure may
be complete or incomplete depending on whether the black hole is
eternal or not. The question is whether the restrictions
(\ref{PP}),(\ref{PPPP}), and (\ref{QQ}) have any significant impact
on the set of all such data.

If we allow the internal spacetime edge to be incomplete, then it is
clear that the kind of argument we used earlier to impose powerful
restrictions on initial value data sets cannot be made to work. For
those restrictions were entirely \emph{global}: they are due to
subtle conditions arising when one tries to extend a solution of an
equation like (\ref{PPP}) to the entire ``initial" surface. If we
allow the Riemannian structure to be incomplete, then we can push
any difficulties which might arise into some region which can then
be conveniently deleted. We conclude that \emph{no black hole which
evaporates entirely can have an independent, internal arrow of
time}. Thus there is no ``reversal" of the arrow inside a ``small"
AdS black hole. This is consistent with the fact that the external
spacetime has its own local arrow [near to but outside the event
horizon] in this case.

We turn now to eternal AdS black holes. In the k = 1 case, we are
dealing with complete Riemannian metrics on manifolds of the form
$\bbr\,\times\,[\m{S}^3/\Gamma]$, where $\Gamma$ is some finite
group [which could be trivial] selected from a known [infinite] list
\cite{kn:wolf}. It can be shown [see \cite{kn:arrow} for a
discussion] that \emph{any} function on S$^3/\Gamma$ can be the
scalar curvature of some metric on that manifold. Even if we confine
our attention to Riemannian products of $\bbr$ with S$^3/\Gamma$,
this already yields a vast number of possible initial metrics, and
of course there will be many more choices if we allow non-product
metrics. The constraints (\ref{PP}), (\ref{PPPP}), and (\ref{QQ})
then place hardly any restrictions on the ``initial" energy density,
energy flux, extrinsic curvature, or spatial metric. [See our
earlier discussion around equation (\ref{PPP}).] We conclude that
there is no internal arrow of time even in the eternal case if the
event horizon is positively curved. [Note that ``positively curved"
here really refers to the \emph{scalar} curvature, so a ``black
ring" event horizon of the form S$^1\,\times\,$S$^2$ is ``positively
curved". Therefore, if black rings exist in the AdS case ---$\,$ see
\cite{kn:empreall} ---$\,$ then their internal thermodynamics can be
expected to be similar to that of AdS black holes with spherical
event horizons.]

The last case is that of eternal AdS black holes with toral event
horizons. Here we need to consider initial data sets
($\m{\rho,\,J^{a},\,K_{ab},\,h_{ab}}$), where now h$_{\m{ab}}$ is a
metric defined on a manifold of topology
$\bbr\,\times\,$[T$^3/\Delta$], where T$^3$ is the three-torus and
where [in the orientable case] $\Delta$ is one of six finite groups
\cite{kn:conway} [including the trivial group]; note again that this
space has to be \emph{complete} with respect to h$_{\m{ab}}$. Now
the manifolds T$^3/\Delta$ have a particular property: they are said
to be \emph{enlargeable} [\cite{kn:lawson}, page 302]. Roughly
speaking this means that they can be ``enlarged" to an arbitrary
extent in all directions by taking a topological covering. This
property of enlargeability [together with the fact that tori and
their orientable non-singular quotients are spin manifolds]
underlies the fact that tori and their quotients are unable to admit
metrics of positive scalar curvature; and this, as we saw above, is
the ultimate reason for the vast reduction in the set of admissible
initial data for toral cosmologies.

Here, however, the situation is more complex, because
$\bbr\,\times\,$[T$^3/\Delta$] is \emph{not} enlargeable. Thus, even
in the case of toral event horizons, it is far from clear that the
argument we used in the cosmological case will work again. Once more
we stress that the [potential] difference between the behaviour of
the arrow of time in the black hole and cosmological cases is not
``built in" \cite{kn:price}: it arises simply because the global
structures of the spatial sections in the two cases are so
different.

Nevertheless, in this particular case we can proceed as follows. We
have to work with the concept of \emph{weak} enlargeability
[\cite{kn:lawson}, page 318]. This is defined by studying certain
maps between Riemannian spaces, which have the effect of changing
the ``sizes" of curvature two-forms rather than that of the space
itself. This is weaker than true enlargeability because
---$\,$ crucially for us
---$\,$ it allows one dimension to escape being enlarged. Because of
this, it turns out that [topological] products of the form
$\bbr\,\times$ E are weakly enlargeable if the space E is
enlargeable. Thus, the spaces $\bbr\,\times\,$[T$^3/\Delta$] are all
weakly enlargeable.

The importance of this is revealed by the following theorem
[\cite{kn:lawson}, page 319], which involves a delicate application
of index theory:

\bigskip
\noindent \textsf{THEOREM (Gromov-Lawson-Kazdan)}: Let
h$_{\m{ab}}$ be a Riemannian metric of non-negative scalar
curvature on a weakly enlargeable manifold, and assume that this
manifold is \emph{complete} with respect to h$_{\m{ab}}$. Then the
Ricci tensor of h$_{\m{ab}}$ must vanish.
\bigskip

Now let us impose the restrictions (\ref{PP}), (\ref{PPPP}), and
(\ref{QQ}) on data sets ($\m{\rho,\,J^{a},\,K_{ab},\,h_{ab}}$)
defined on a spacetime edge of the form
$\bbr\,\times\,$[T$^3/\Delta$]. By the Gromov-Lawson-Kazdan theorem,
we see that the left side of (\ref{PP}) has to vanish identically;
hence, so must each term on the right, and so, by (\ref{PPPP}), must
J$^{\m{a}}$. Thus we drastically cut down the number of possible
data sets: they must be of the form ($0,\,0,\,0,\,$ q$_{\m{ab}}$),
where q$_{\m{ab}}$ is a complete metric which is very severely
constrained by the requirement that its Ricci tensor vanishes
exactly.

In fact, with specific spatial boundary conditions, this restriction
will essentially determine q$_{\m{ab}}$ \emph{uniquely}. But since
the spatial sections extend to infinity here, the field theory at
infinity will supply such boundary conditions. Thus q$_{\m{ab}}$ is
fixed, and once again we have essentially unique initial data in
this case. With the boundary conditions one expects here,
q$_{\m{ab}}$ will be highly symmetric; see the discussion of the
Lemma stated in Section 5 below. It is certainly unique [apart from
well-understood ambiguities of the kind to be discussed in the next
section], and maximally locally symmetric, for
\emph{four-dimensional} toral AdS black holes. For the spatial
sections will then be three-dimensional, and it is an elementary
fact that Ricci-flat metrics in that case are exactly flat. [In four
dimensions there are of course non-flat Ricci-flat metrics on
topologically complicated manifolds such as K3. But the universal
cover of $\bbr\,\times\,$[T$^3/\Delta$] is just the topologically
trivial space $\bbr^4$, and any metric on
$\bbr\,\times\,$[T$^3/\Delta$] pulls back to $\bbr^4$; with the kind
of boundary conditions we expect here, it seems very likely that the
spatial metric will have to be flat ---$\,$ given that it is
\emph{complete} ---$\,$ in this case also.]

We see that there is an essentially unique ``initial" data set for
the interiors of toral AdS black holes, just as there is for toral
cosmology, despite the non-compactness of the sections in the
present case. We conclude that these AdS black holes, \emph{and
these only}, do have an internal arrow of time pointing in the
opposite direction to what one would expect; that is, their arrow of
time is reversed relative to their counterparts with positively
curved event horizons. We note in passing that this part of the
theory contradicts the Weyl Curvature Hypothesis \cite{kn:penrose}
[which states that the Weyl curvature should vanish \emph{only} at
the ``initial" cosmological spacetime edge].

We see that some terms in the sum over geometries do have unusual
internal thermodynamic behaviour, while others do not. Let us turn
to a more precise survey of all of the terms in this sum.

\addtocounter{section}{1}
\section* {\large{\textsf{5. Different Black Holes with the Same Entropy}}}
The next step is to perform the ``sum over geometries" [analogous to
the ``sums" discussed by Maldacena \cite{kn:juan} and Hawking
\cite{kn:hawk}] to see whether the presence of toral black holes in
the sum leads to the appropriate modification of the
Horowitz-Maldacena proposal. Unfortunately, it is completely unknown
how this can be done at this level of detail. We can however specify
precisely the full set of AdS black holes over which the sum will
have to be performed. We can also point out some hints from global
geometry as to which terms in the sum are likely to dominate.

A realistic non-rotating neutral black hole in de Sitter spacetime
is determined by its entropy. For there are no topological
ambiguities in this case: a black hole formed by the collapse of a
star obviously has a topologically spherical event horizon, and its
mass is fixed by the area of this horizon.

The [five-dimensional] AdS black holes we have been discussing in
this work, by contrast, are by no means specified by their entropy.
There are several levels of ambiguity. First, given any value for
the entropy, there will be AdS black holes of the same entropy with
either positively curved or flat event horizons. Then, within each
class, there are still many different black holes with that entropy.
Let us consider the positive case first.

The entropy of the hole with metric (\ref{A}) is proportional to the
area of the event horizon, which is given by
\begin{equation}\label{S}
\m{A_{eh}\;=\;A_k\,r_{\m{eh}}^3},
\end{equation}
where r$_{\m{eh}}$ is the value of the radial coordinate at that
surface. If k = 1, A$_{\m{eh}}$ is given by
\begin{equation}\label{T}
\m{A_{eh}\;=\;2^{\,-3/2}\,L^3\,A_1\,\Bigg[-
1\;+\;\Bigg\{1\;+\;{64\pi M\over 3A_1L^2}\Bigg\}^{1/2}\Bigg]^{3/2}}.
\end{equation}
For such a black hole, A$_1$ is 2$\pi^2$ if the topology of the
event horizon is that of the [unit] sphere S$^3$. But as we know,
the event horizon can have the topology of any non-singular quotient
of the form S$^3/\Gamma$, where $\Gamma$ is a finite group
\cite{kn:wolf}. For example, $\Gamma$ can be any of the well-known
ADE finite subgroups of SU(2) [the cyclic, quaternionic, and binary
polyhedral groups], but there are other possibilities. If $|\Gamma|$
is the order of this group, then the value of A$_1$ for S$^3/\Gamma$
is 2$\pi^2/|\Gamma|$. Thus A$_1$ can have many values extending
downwards from 2$\pi^2$; because $|\Gamma|$ is arbitrarily large for
the lens spaces S$^3/\bbz_n$, where $\bbz_n$ is the cyclic group of
order n, A$_1$ can in fact be arbitrarily small. It takes the form
2$\pi^2$/n, where every integer n is possible.

If we think of the area of the event horizon as a function of A$_1$,
as given in equation (\ref{T}), then it has an interesting behaviour
as A$_1$ decreases from its maximum value of 2$\pi^2$. If the mass
of the black hole is small compared to L$^2$ [that is, if it has a
value typical for a black hole which evaporates completely], then
taking the quotient of the event horizon by small groups $\Gamma$
actually \emph{increases} the entropy of the black hole for a fixed
value of M. For larger groups, however, the entropy is decreased by
taking the quotient. For [sufficiently] ``large" black holes
[meaning that M is large compared to L$^2$], the effect of taking
the quotient is always to decrease the entropy.

More importantly for our purposes: suppose that we fix the entropy,
that is, we fix the area of the event horizon. Then the
dimensionless quantity M/L$^2$ can be expressed in terms of A$_1$:
\begin{equation}\label{U}
\m{M/L^2\;=\;{3\over 16\pi}\Bigg({A_{eh}^{2/3}\over
L^2}\Bigg)\Bigg[\Bigg({A_{eh}^{2/3}\over
L^2}\Bigg)\,A^{-\,1/3}_1\;+\;A^{1/3}_1 \Bigg].}
\end{equation}
The fact that M can always be found for any value of A$_1$ means
that, if A$_1$ changes [by taking quotients as above], the effects
of this can always be compensated by choosing M appropriately. The
function on the right side in this equation has a global minimum at
a formal value of A$_1$ = A$_{\m{eh}}$/L$^3$, so if
A$_{\m{eh}}$/L$^3$ is much smaller than 2$\pi^2$ then for a finite
number of steps downwards one has to reduce M in order to keep the
entropy constant; subsequently [and always if A$_{\m{eh}}$/L$^3$ is
not small] M has to be increased.

We see from this that \emph{for each specified value of the entropy,
there is a countable infinity of AdS black holes with positively
curved event horizons having that entropy.} Note that these black
holes are not uniquely fixed by their masses [or by the integer
$|\Gamma|$], because two different AdS black holes with positively
curved event horizons can have the same entropy and the same mass:
consider for example the black holes of mass M and with event
horizons having the respective geometries of S$^3/\bbz_{120}$,
S$^3$/Q$_{120}$, and S$^3$/$\tilde{\m{I}}$, where Q$_{120}$ is the
quaternionic [binary dihedral] group of order 120 and
$\tilde{\m{I}}$ is the binary icosahedral group \cite{kn:wolf}. In
all three cases we have A$_1$ = $\pi^2$/60, so all three holes have
the same entropy.

The sum over this sector of AdS black holes therefore takes the form
of a discrete ``sum" over the candidates for $\Gamma$, with groups
of larger order generally corresponding to larger masses. We stress
that the candidates for $\Gamma$ are completely and explicitly
known: see \cite{kn:wolf}.

We now turn to the case of AdS black holes with flat event horizons.
Here the situation is quite different: there is a limited range of
possible topologies, but many continuous parameters. The possible
topologies \cite{kn:conway}\cite{kn:flatweeks} in the orientable
case [to which we shall confine ourselves in this work] are just the
torus or torocosm T\3, the dicosm T\3/$\bbz_2$, the tricosm
T\3/$\bbz_3$, the tetracosm T\3/$\bbz_4$, the hexacosm T\3/$\bbz_6$,
and the didicosm or Hantzsche-Wendt space
T$^3$/[$\bbz_2\;\times\;\bbz_2$]. In each case, there are continuous
parameters which distinguish manifolds of the same topology which
have different [global] geometries. For example, in the case of the
torus, one can cover $\bbr^3$ with tiles of various shapes and
sizes; identifying all of the tiles produces, in every case, a space
with the topology of T$^3$. There are six continuous parameters
describing the possible shape and size of the fundamental domain for
a torus. For the other orientable topologies, the number of
continuous parameters is smaller [because some parameters have to be
fixed in order to perform the projection to the quotient]. There are
four for T\3/$\bbz_2$, two for T\3/$\bbz_3$, two for T\3/$\bbz_4$,
two for T\3/$\bbz_6$, and three for
T$^3$/[$\bbz_2\;\times\;\bbz_2$]; see \cite{kn:conway} for the
details.

In each case, one can think of the parameter K, used in Section 3,
as a function of the continuous parameters; in each case, K ranges
continuously from arbitrarily small to arbitrarily large values. If
we fix the entropy of the black hole, then from equation (\ref{F})
we have
\begin{equation}\label{V}
\m{M/L^2\;=\;{3\over 32\pi^2}\Bigg({A_{eh}^{4/3}\over
L^4}\Bigg)\,K^{-1}.}
\end{equation}
Clearly it is always possible to adjust M to keep A$_{\m{eh}}$
fixed, no matter how K varies over the six sets of flat compact
orientable three-manifolds; for a given A$_{\m{eh}}$, pick any of
the six topologies, choose the parameters, compute K, and use
(\ref{V}) to deduce M. In other words, \emph{for each specified
value of the entropy, there is an uncountable infinity of AdS black
holes with flat event horizons having that entropy.} The ``sum" over
black holes in this case is a discrete sum over the six topological
classes, followed by a continuous sum over the size and shape
parameters.

In conclusion, then, we have a complicated but complete and explicit
procedure for finding all of the stable AdS black holes with a
specified value for the entropy: we have two choices for the overall
type of hole, discrete choices of the detailed topology in each
class, and, in one case, a choice of continuous parameters. The
problem now is to understand how to weight the terms in the ``sum"
and how to evaluate it.

There is an obvious sense in which, for fixed entropy, the black
holes with flat event horizons ``outnumber" those with positively
curved event horizons. This is where the question of weighting the
terms in the ``sum" is crucial. The following line of thought may be
relevant.

Our discussion in Section 3 underlined the fact that black holes
with flat event horizons are, in a sense, ``closer to being
unstable" than their counterparts with positively curved event
horizons. One can in fact use the global techniques we have been
discussing here to formulate another sense in which these holes are
``close to instability": one can prove, using the theorem of Schoen
et al stated above [and other methods discussed in \cite{kn:arrow}],
the following lemma.

\bigskip
\noindent \textsf{LEMMA}: Let g be a Riemannian metric on a manifold
with the topology of a torus or of a non-singular quotient of a
torus. Then unless g is conformal to an exactly flat metric, it is
conformal to a metric of constant negative scalar curvature.
\bigskip

The Euclidean version of the metric (\ref{A}) for k = 0 is defined
on a space which, after Euclidean time is compactified, does have a
conformal boundary with the topology of a compact flat manifold; and
the conformal structure at infinity is indeed that of a perfectly
flat metric. But the Lemma means that any non-conformal deformation
of the geometry at infinity will result in negative scalar curvature
there. As we discussed in Section 2, this would immediately lead to
an instability of the kind discussed by Seiberg and Witten. Thus the
flat case is indeed ``close" to being unstable in this strong
[global-geometric] sense. Possibly this reduces the weight of black
holes with flat event horizons in the ``sum" over AdS black holes,
so that they only make a small [but crucial] contribution to a sum
which is dominated by spherical holes. [For example, if in general
the boundary can be ``set free" in the AdS/CFT correspondence
\cite{kn:compere}, the above Lemma apparently implies that this is
possible for spherical black holes but not for toral ones. Note that
this ``rigidity" enforces the geometrically regular boundary
conditions discussed above when we applied the Gromov-Lawson-Kazdan
theorem.]

One might also wish eventually to consider whether ``black rings"
---$\,$ \emph{if} they exist in the asymptotically AdS case
\cite{kn:empreall} ---$\,$ may also have to be included; also
whether unusual behaviour of the kind described in
\cite{kn:odintsov} may have to be taken into account, along with
recent, more sophisticated analyses of the nature of Hawking
radiation itself \cite{kn:vach}. A deeper understanding of the
mechanism which prevents the black hole spacetime edge from being
singular will of course be necessary; see for example \cite{kn:stoj}
for a recent discussion of one aspect of this issue. Finally, one
would of course like to relate any string-theoretic account of black
hole evaporation to the well-known analysis of black hole entropy
given by Strominger and Vafa \cite{kn:vafastrom}.

\addtocounter{section}{1}
\section* {\large{\textsf{6. Conclusion: The Centrality of the Arrow}}}

The black hole information loss problem is disconcerting, because it
apparently requires us to believe that we have failed to understand
some fundamental aspect of either quantum mechanics or general
relativity at scales where both are exceedingly well-tested. The
Horowitz-Maldacena proposal \cite{kn:horomald} offers an escape from
this dilemma. In that sense it is, despite its strangeness, the most
conservative approach to the problem, and as such it merits further
attention.

One of the main objectives of this work is to point out that the
objection made by Gottesman and Preskill \cite{kn:preskill} to this
proposal is essentially thermodynamic in nature. It is therefore
based on the assumption that we understand the [gravitational]
thermodynamics of systems near to spacetime edges. But in at least
one case
---$\,$ the spacetime edge associated with the beginning of our
Universe ---$\,$ this understanding is work in progress: the origin
of the arrow of time has long been and remains a matter of much
debate. \emph{It follows that we need to agree on a theory of the
arrow before we can decide whether the Gottesman-Preskill objection
is fatal to the Horowitz-Maldacena proposal.} Conversely, the
ability to salvage this proposal would be strong evidence in favour
of any particular theory of the arrow.

In this work we have outlined some ideas leading in this
direction. First, we note that both Maldacena \cite{kn:juan} and
Hawking \cite{kn:hawk} have emphasised that the information loss
problem cannot be solved by studying just one geometry: some kind
of as yet poorly understood ``sum over geometries" will be
required. Next, we observe that the correspondence between
five-dimensional de Sitter black holes and their AdS$_5$
counterparts [for which there is strong evidence that unitarity is
indeed preserved, particularly in the eternal case] is far from
being one-to-one; the range of possible AdS$_5$ holes with a given
entropy is large but can be surveyed. One therefore knows exactly
which geometries will have to be ``summed over". Crucially, we
applied a specific theory of the arrow to deduce that the internal
thermodynamics of AdS black holes is ``normal" for spherical holes
but ``reversed" for toral holes. A sum over all AdS$_5$ black
holes of the same entropy will therefore sample thermodynamic
behaviour of both kinds; and there are hints that the sum will be
dominated by the ``normal" variety, though not to the total
exclusion of the more unusual behaviour found in the toral case.
This holds out hope that the Gottesman-Preskill objection can
eventually be answered.

\addtocounter{section}{1}
\section*{\large{\textsf{Acknowledgement}}}
The author is grateful to Dr. Soon Wanmei for useful discussions.

\end{document}